\documentclass[twocolumn,prl]{revtex4-1}
\usepackage{xcolor}
\usepackage{pdfcolmk}
\usepackage{amsmath}
\usepackage{amssymb}
\usepackage{graphicx}
\usepackage{epstopdf}
\PassOptionsToPackage{normalem}{ulem}
\usepackage{ulem}
\usepackage[unicode=true,pdfusetitle,
 bookmarks=true,bookmarksnumbered=false,bookmarksopen=false,
 breaklinks=false,pdfborder={0 0 0},pdfborderstyle={},backref=false,colorlinks=true]
 {hyperref}
\hypersetup{
 colorlinks,linkcolor=red,citecolor=blue}

\makeatletter

\providecolor{lyxadded}{rgb}{0,0,1}
\providecolor{lyxdeleted}{rgb}{1,0,0}

\DeclareRobustCommand{\lyxsout}[1]{\ifx\\#1\else\sout{#1}\fi}

\usepackage{amsfonts}

\makeatother

\begin{document}

\title{Magnonic frequency comb in the magnomechanical resonator}

\author{Guan-Ting Xu$^{1,2,3}$}
\thanks{These authors contributed equally to this work.}
\author{Mai Zhang$^{1,2,3}$}
\thanks{These authors contributed equally to this work.}
\author{Yu Wang$^{1,2}$}
\author{Zhen Shen$^{1,2,3}$}
\email{shenzhen@ustc.edu.cn}
\author{Guang-Can Guo$^{1,2,3}$}
\author{Chun-Hua Dong$^{1,2,3}$}
\email{chunhua@ustc.edu.cn}

\affiliation{$^{1}$CAS Key Laboratory of Quantum Information, University of Science
and Technology of China, Hefei, Anhui 230026, People's Republic of
China}

\affiliation{$^{2}$CAS Center For Excellence in Quantum Information and Quantum
Physics, University of Science and Technology of China, Hefei, Anhui
230088, People's Republic of China}

\affiliation{$^{3}$Hefei National Laboratory, University of Science and Technology
of China, Hefei, Anhui 230088, People\textquoteright s Republic of
China}

\begin{abstract}
An optical frequency comb is a spectrum of optical radiation which
consists of evenly spaced and phase-coherent narrow spectral lines
 and is initially invented in laser for frequency metrology purposes.
A direct analogue of frequency combs in the magnonic systems has not
been demonstrated to date. In our experiment, we generate a new magnonic
frequency comb  in the resonator with giant mechanical oscillation
through the magnomechanical interaction. We observe the magnonic
frequency comb contains up to 20 comb lines, which are separated
to the mechanical frequency of the $10.08\ \mathrm{MHz}$. The thermal
effect based on the strong pump power induces the cyclic oscillation
of the magnon frequency shift, which leads to a periodic oscillation
of the magnonic frequency comb. Moreover, we demonstrate the stabilization
and control of the frequency spacing of the magnonic frequency comb
via injection locking. Our work lays the groundwork of magnonic frequency
combs for sensing and metrology.
\end{abstract}
\maketitle

\subsection{Introduction}

Optical frequency combs are composed of a set of equidistant coherent
optical lines in the frequency domain. In the last two decades, they
have exhibited rapid development in various fields \cite{kippenberg2011microresonator,kippenberg2018dissipative,diddams2020optical},
such as astronomy and cosmology, optical atomic clock \cite{newman2019architecture},
light detection and ranging (LiDAR) \cite{suh2018soliton,trocha2018ultrafast,wang2020long},
low-noise microwave source \cite{liang2015high,lucas2020ultralow,liu2020photonic},
coherent optical communication \cite{marin2017microresonator,corcoran2020ultra},
quantum key distribution \cite{wang2020quantum}, dual-comb spectroscopy
\cite{suh2016microresonator,yu2018silicon}, spectrometer \cite{yang2019vernier,niu2023khz},
and optical coherence tomography \cite{ji2019chip,marchand2021soliton}.
Optical frequency combs in microresonators can be generated through
electro-optical modulation or third-order Kerr nonlinearity $\chi^{(3)}$
\cite{kippenberg2018dissipative}. Recently, mechanical vibration
has been demonstrated for the generation and engineering of optical
combs \cite{cao2014phononic,de2023mechanical,ganesan2017phononic,miri2018optomechanical,hu2021generation},
which has attracted increasing attention due to its low repetition
rate and suitability for acoustic sensing \cite{zhang2021optomechanical}.
Furthermore, the frequency combs have also been well investigated
beyond optical systems, such as microwave system and phononic system.
However, the frequency comb in the magnonic system has not been demonstrated
to date.

Magnons are the quantum of collective spin excitation of magnetization
in ferromagnetic insulators such as yttrium iron garnet (YIG). They
exhibit great frequency tunability, extremely low magnetic damping
and high Curie temperature, making them an ideal carrier for performing
coherent information processing, and precision measurements \cite{serga2010yig,lenk2011building,liu2019phase,wang2020chiral,chai2022single,shen2022coherent}.
Similar to the traditional frequency comb, the magnonic frequency
comb has been proposed for the development of high-precision magnonic
frequency metrology and spectroscopy via the nonlinear interaction
\cite{rao2023unveiling,wang2021magnonic,wang2022twisted}. However,
the weak nonlinear interaction of magnons presents a challenge for
generating magnonic frequency combs \cite{liu2022optomagnonic}. Given
the successful of introduction mechanical vibration, the magnonic
frequency comb based on magneto-mechanical coupling has been proposed
in theory \cite{xiong2023magnonic}.

In this work, we experimentally generate a magnonic frequency comb
in the resonator with giant mechanical oscillation through the magnomechanical
interaction. This dynamical process is facilitated by an external
pump that mimicked magnomechanical interactions mediated through the
magnetostrictive effect. When the pump power is strong enough, magnomechanical
nonlinearities play a significant role, and a self-induced nonlinear
phenomenon similar to the Kerr-frequency comb effect can be observed
in the magnomechanical system. We observe a magnonic frequency comb
with up to 20 comb lines and a frequency spacing of $10.08\ \mathrm{MHz}$,
equaling the resonant frequency of the mechanical resonator. In this
process, the strong pump field on the magnon is required, leading
to a raised temperature of the magnon and a corresponding magnon frequency
shift. Then, the magnonic frequency comb shows periodic oscillation
due to the thermal effect. It is also in good agreement with the
theoretical calculation based on the evolution equations of the magnon
and phonon modes. Finally, we demonstrate the stabilization and control
of the frequency spacing of the magnonic frequency comb via injection
locking \cite{wan2020frequency}. We can achieve tuning of comb teeth
beyond a range of $1\ \mathrm{kHz}$.

\subsection{Experimental Results}

\begin{figure}
\centering{}\includegraphics[clip,width=1\columnwidth]{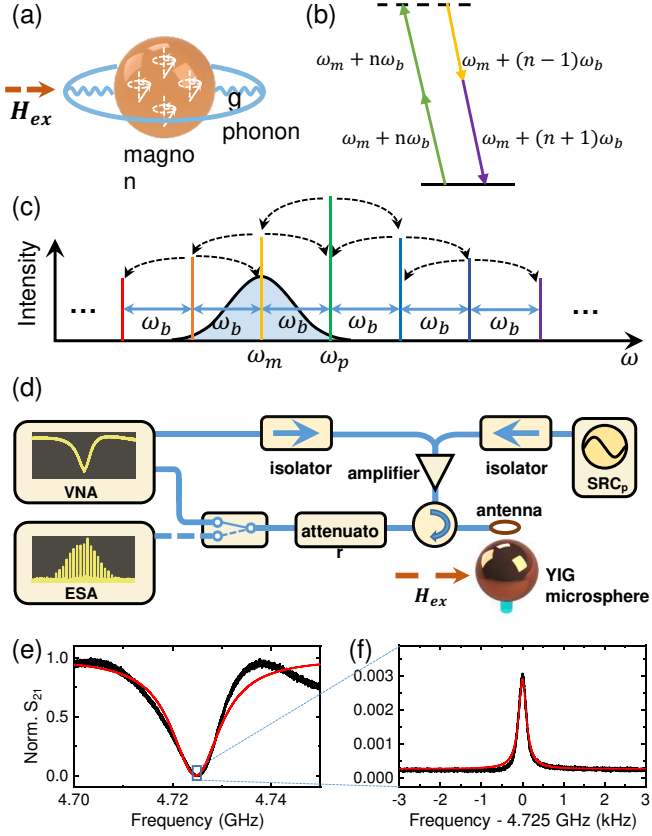}\caption{\label{fig:Fig1}(a) Illustration of the coupling between magnon and
phonon with single magnon-phonon coupling strength $g$. (b) The magnonic
frequency comb is generated via a two-magnon process. Two magnons
with frequency $\omega_{m}+n\omega_{b}$ are annihilated to create
two magnons with frequency $\omega_{m}+(n-1)\omega_{b}$ and $\omega_{m}+(n+1)\omega_{b}$,
where the $\omega_{m}$ and $\omega_{b}$ are the frequency of the
magnon and mechanical modes, respectively. (c) A cascaded two-magnon
process is initiated to form a frequency comb separated with a mechanical
frequency of $\omega_{b}$. (d) Schematic of the experimental setup
for the magnonic frequency comb. SRC: microwave source; VNA: vector
network analyzer; ESA: electrical spectrum analyzer. (e) Measured
microwave reflection spectrum around $4.725\:\mathrm{GHz}$. (f) The detailed
spectrum showing the mechanical mode in (e). The red lines are results
of theoretical calculations discussed in the main text with the parameters
of $g=2.36\:\mathrm{mHz}$, $\kappa_{m}=12\:\mathrm{MHz}$, $\kappa_{b}/2\pi=220\mathrm{\:Hz}$
and $\kappa_{in}=6.1\mathrm{\:MHz}$.}
\end{figure}

Due to magnetostrictive forces, a type of radiation pressure-like
interaction arises between the magnon mode and the mechanical mode
, as shown in Fig. 1(a). When increasing the pump power to enhance
the interaction, the self-induced nonlinear phenomena similar to
the Kerr-frequency comb effect can be observed in the magnomechanical
system based on the cascaded two-magnon process, as illustrated in
Fig. 1(b-c). The two magnon process including two magnons with frequency
of $\omega_{m}+n\omega_{b}$ are annihilated to create two magnons
with frequency $\omega_{m}+(n-1)\omega_{b}$ and $\omega_{m}+(n+1)\omega_{b}$,
where the $\omega_{m}$ and $\omega_{b}$ are the frequency of the
magnon and mechanical modes, respectively. This initiates a cascaded
process leading to the formation of a frequency comb with a free spectral
range (FSR) of $\omega_{b}$, the frequency relationship between the
magnon frequency $\omega_{m}$, mechanical frequency $\omega_{b}$
and pump frequency $\omega_{p}$ is given by $\omega_{m}=\omega_{p}-\omega_{b}$.
The interaction between the magnon and mechanical modes can be represented
by the Hamiltonian expression:

\begin{equation}
\begin{array}{cc}
H= & (\omega_{m}-\omega_{p}-i\frac{\kappa_{m}}{2})m^{\dagger}m+(\omega_{b}-i\frac{\kappa_{b}}{2})b^{\dagger}b\\
 & +g\left(b^{\dagger}+b\right)m^{\dagger}m+i\sqrt{\kappa_{in}}\varepsilon_{p}\left(m^{\dagger}-m\right)
\end{array}
\end{equation}
where $m$ and $b$ are the annihilation operators of the magnon and
mechanical modes, $\kappa_{m}$ and $\kappa_{b}$ represent the dissipation
of magnon and phonon, respectively. $g$ is the single magnon-phonon
coupling strength, which depends on the overlap between magnon and
phonon modes. $\varepsilon_{p}$ is the pump field of the microwave
with the frequency of $\omega_{p}$. $\kappa_{in}$ is the input coupling
rate of the microwave. For the blue-detuned pump, the relationship
between $\omega_{m}$ and $\omega_{b}$ satisfies $\Delta=\omega_{p}-\omega_{m}=\omega_{b}$.

As shown in the Hamiltonian, coupling between magnons and phonons
leads to the phonon number dependent magnon frequency shift:
\begin{eqnarray}
\frac{dm}{dt} & = & \{i[-\Delta-g(b^{\dagger}+b)]-\frac{\kappa_{m}}{2}\}m+\sqrt{\kappa_{in}}\varepsilon_{p},\\
\frac{db}{dt} & = & (i\omega_{b}-\frac{\kappa_{b}}{2})b-igm^{\dagger}m.
\end{eqnarray}
Like the frequency comb induced by the Kerr oscillator, when the pump
condition exceeds the threshold of intra-cavity field instability,
the mechanical mode starts to lasing. Modulated by the strongly lasing
mechanical mode, magnon comb generates, where the amplitude of magnon
mode will take a form as $m(t)=\sum_{n}m_{n}e^{i(\omega_{m}-\omega_{p})t}e^{-in\omega_{b}t},n\in Z$,
where $m_{n}=\sqrt{\kappa_{in}}\varepsilon_{p}\sum_{k}\frac{J_{k-n}(\xi)J_{n}(-\xi)}{\kappa_{m}/2-i(\Delta+k\omega_{b})}$
and $\xi=2gb/\omega_{b}$ is the normalized amplitude of mechanical
mode. Meanwhile, the linewidth of mechanical mode is also narrowed
significantly by the magnon comb that $\kappa_{b}^{\prime}=\kappa_{b}-\chi$.
As the narrow linewidth of the mechanical mode, the no-coherence part
of the magnon pump can be ignored and there is $\chi=2g\sum_{n}\mathrm{Im[}m_{n}^{\dagger}m_{n+1}]/b$.
When $\kappa_{b}^{\prime}\leq0$, the system is unstable and a magnon
comb occurs.

\begin{figure*}
\begin{centering}
\includegraphics[clip,width=1.8\columnwidth]{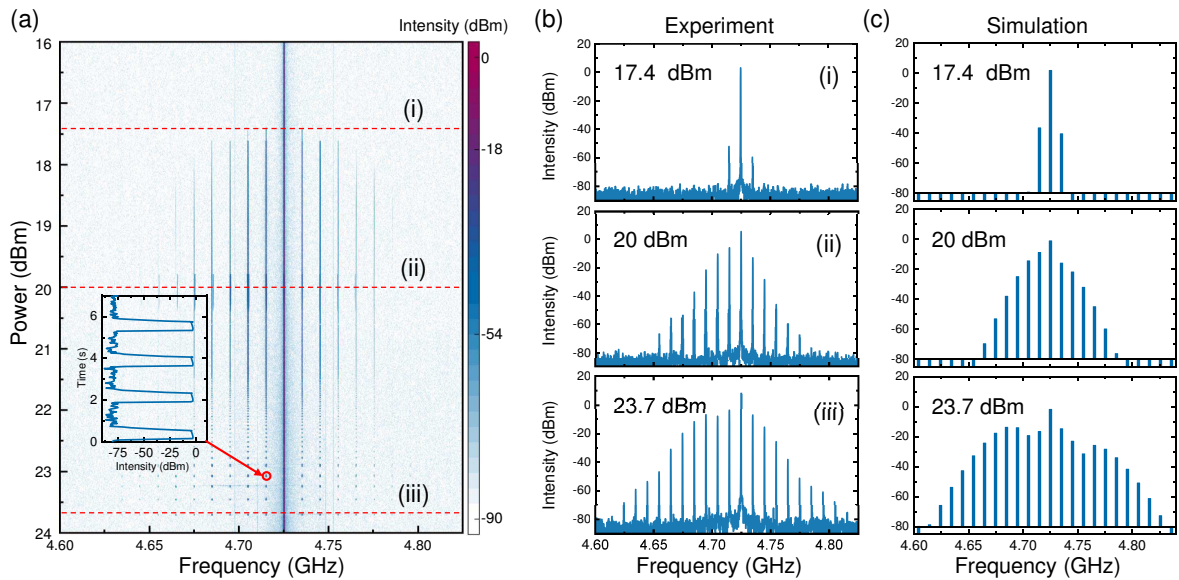}
\par\end{centering}
\caption{\label{fig:Fig2} (a) Evolution of the magnonic frequency comb along
increased pump power from $16\:\mathrm{dBm}$ to $24\:\mathrm{dBm}$
with the frequency of $\omega_{p}=4.725\:\mathrm{GHz}$. The inset
shows the evolution of the spectrum over a period of 7 seconds at
a power of $23.1\:\mathrm{dBm}$ and frequency $\omega_{p}-\omega_{b}=4.714\:\mathrm{GHz}$
component. (b) Snapshots with different evolution stages in (a). (c)
The simulated spectra of the magnonic frequency combs with the same
conditions in (b). }
\end{figure*}
The experiment setup is illustrated in Fig. 1(d). Initially, a YIG
microsphere with a diameter of $623.6\:\mathrm{\mu m}$ is subject
to a bias magnetic field $H_{ex}$ aligned parallel to the equatorial
plane of the YIG microsphere, which supports a uniform magnon mode.
The relation between the frequency of the magnon and magnetic field
intensity follows the equation $\omega_{m}=\gamma H_{ex}$, where
$\gamma=2\pi\times2.8\:\mathrm{MHz/Oe}$ represents the gyromagnetic
ratio. In order to excite the magnon mode, we employ an antenna located
close to the YIG microsphere with a frequency around $4.725\:\mathrm{GHz}$.
Fig. 1(e) shows the magnon resonance through the microwave reflection
spectrum $S_{21}$ obtained by a vector network analyzer (VNA), corresponding
to a dissipation rate of $\kappa_{m}=12\:\mathrm{MHz}$. The frequency
associated with the mechanical mode is obtained $\omega_{b}/2\pi=10.08\mathrm{\:MHz}$ \cite{shen2022coherent,xu2023ringing}.
The dissipation rate of the mechanical mode is $\kappa_{b}/2\pi=220\mathrm{\:Hz}$, as shown
in Fig.1(f). Meanwhile the
pump field of the microwave with the frequency of $\omega_{p}=4.725\:\mathrm{GHz}$.
Based on these interactions, the theoretical calculations are achieved
with the parameter of the single magnon-phonon coupling strength $g=2.36\:\mathrm{mHz}$
and the input coupling rate of microwave $\kappa_{in}=6.1\mathrm{\:MHz}$.

Figure 2 displays the spectra of the magnonic frequency comb with
varying pump power from $16\:\mathrm{dBm}$ to $24\:\mathrm{dBm}$
at the frequency of $\omega_{p}=4.725\:\mathrm{GHz}$. For a weak
microwave pump field, i. e., below the threshold of $17.4\:\mathrm{dBm}$,
the spectrum only contains the pump component, with no comb lines.
As the pump power gradually increases to reach the threshold for generating
comb conditions, a frequency comb emerges, as shown in Fig. 2(b),
with a tooth spacing of $\omega_{b}=10.08\:\mathrm{MHz}$. The number
of comb lines increases steadily with higher pump power. The typical
magnonic frequency comb can have up to 21 comb lines with the pump
power at $23.7\:\mathrm{dBm}$. Figure 2(c) displays magnonic frequency
combs obtained from numerical simulations under the pump power of
$17.4\:\mathrm{dBm}$, $20\:\mathrm{dBm}$, and $23.6\:\mathrm{dBm}$,
respectively. Since strong pump heats the system, magnon mode is red shifted due to the thermal effect and the effective detuning between pump frequency and magnon
frequency $\Delta\omega_{eff}$ are $1.1\omega_{b}$, $1.3\omega_{b}$
and $2\omega_{b}$, respectively at those pump powers above. These
inconsistent detunings are mainly caused by thermal effects, which
we will discuss later. Other numerical simulation parameters are consistent
with the experiment.

During the process of increasing pump power, it is observed that when
the pump power reached $21.4\:\mathrm{dBm}$, the comb teeth become
unstable and oscillate in the time domain, with the oscillation period
increasing as the power increases. The inset of Fig. 2(a) shows the
evolution of a typical comb line over a period of $7$ seconds at
a power of $23.1\:\mathrm{dBm}$ at a frequency of $\omega_{p}-\omega_{b}=4.714\:\mathrm{GHz}$.
To investigate the physical mechanism for this oscillation, we further
investigate the dynamic evolution of the magnon mode. When we fix
the pump frequency at $\omega_{p}=4.725\:\mathrm{GHz}$ and the power
at $23.1\:\mathrm{dBm}$, we observe the temporal evolution of the
$S_{21}$ spectrum measured by a vector network analyzer (VNA), as shown
in Fig. 3(a). The comb generation exhibits a periodicity of $T=1.7\:\mathrm{s}$,
corresponding to the period shown in the inset of Fig. 2(a). Figure
3(b-c) show the typical $S_{21}$ spectrum at $t=2\:\mathrm{s}$ and
$3\:\mathrm{s}$ in Fig. 3(a), respectively. When the pump power is
very strong, an artifact signal peak will appear at the overlap of
the comb tooth frequency $(4.714\:\mathrm{GHz})$ in the $S_{21}$ spectrum,
and its frequency is related to the resolution bandwidth of the VNA.
In addition, the oscillation time scale is similar to that of thermal
relaxation \cite{carmon2004dynamical,hu2021generation}. We conduct
further analysis to investigate the thermal effect during tooth generation.
With a large amount of energy coupled into the system, the YIG sphere
will be heated:
\begin{equation}
\frac{d\delta T}{dt}=-\frac{1}{\tau}\delta T+\frac{\kappa_{in}\hbar\omega_{p}m^{\dagger}m}{c_{p}},
\end{equation}
where $\delta T$ is the temperature difference between YIG and environment,
$\tau=1.6\:\mathrm{s}$ is the thermal relaxation time, $c_{p}=9.55\times10^{-3}\:\mathrm{J/K}$
is the thermal capacity of the YIG sphere and $\kappa_{in}=\kappa_{m}-\kappa_{ex}$
is the intrinsic dissipation rate. As a result, the frequency of the
magnon mode will be red shifted with the increasing temperature:
\begin{equation}
\omega_{m}^{\prime}=\omega_{m}-\alpha_{\mathrm{T}}(\delta T-\delta T_{0}),
\end{equation}
where $\alpha_{\mathrm{T}}=0.41\:\mathrm{MHz/K}$ is the frequency
shift per unit temperature change and $\delta T_{0}=4.2\:\mathrm{K}$
is the initially temperature difference. Figure 3(d) depicts the threshold
curve of phonon amplitude versus detuning, where $\kappa_{b}^{\prime}=0$.
Points on the curve indicate steady states. For states under the curve
corresponding to the shaded fill area in Fig. 3(d), there is $\kappa_{b}^{\prime}<0$
and phonon amplitude will increase. Conversely, for states above the
curve corresponding to the blank space in Fig. 3(d), there is $\kappa_{b}^{\prime}>0$
and the phonon amplitude will decrease. Moreover, as shown in Fig.
3(d), this process can be divided into four processes: (1) When the
threshold is approached, a large number of magnons are generated,
causing the YIG sphere to heat up and leading to a red shift in the
magnon frequency due to thermal effects. (2) When the frequency shift
is substantial enough, the threshold cannot be approached for all
amplitudes of the mechanical mode. At this time, the magnon number
will dramatically decrease due to the large detuning between resonances.
(3) The YIG sphere will no longer heat up, and the heat will be dissipated
into the ambient air, causing the magnon frequency to shift back in
the direction of high frequency. (4) If the pump power remains constant,
the magnon frequency will return to the point where the threshold
can be reached, and comb teeth are generated. Then the state of the
system comes back to process (1) and the oscillation will repeat steadily.
We examine the frequency shift and oscillation period change under
different pump powers, as shown in Fig. 3(e-f). We find that the frequency
shift varied from $1\:\mathrm{MHz}$ to $7\:\mathrm{MHz}$, while
the oscillation period increased from $0.4\:\mathrm{s}$ to $1.7\:\mathrm{s}$
with increasing pump power. This phenomenon is due to the different
heating rates of the YIG sphere with varying pump power, which agrees
well with the numerical calculations.
\begin{figure}
\begin{centering}
\includegraphics[clip,width=1\columnwidth]{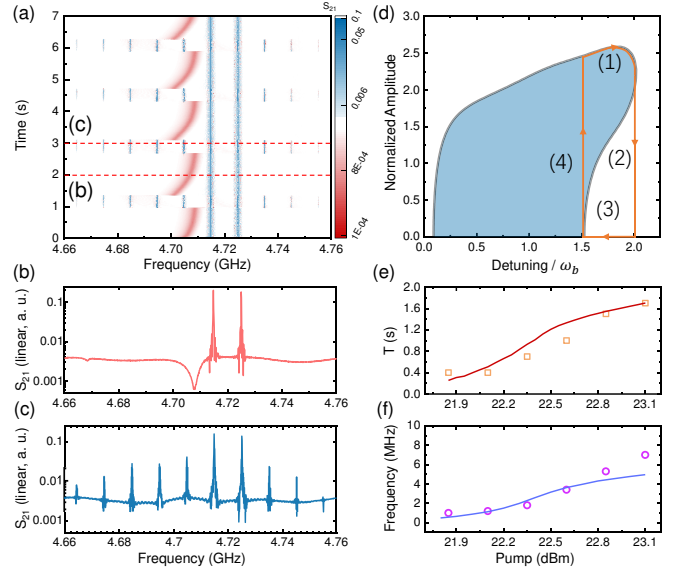}
\par\end{centering}
\caption{\label{fig:Fig3}(a) Evolution of the $S_{21}$ spectrum along time
(within $7$ seconds) with the pump power of $23.1\:\mathrm{dBm}$
and frequency of $\omega_{p}=4.725\:\mathrm{GHz}$. (b-c) The typical
$S_{21}$ spectrum under $t=2\:\mathrm{s}$ and $3\:\mathrm{s}$ in
(a). (d) Threshold curve for phonon amplitude and detuning at the
pump power of $23.1\:\mathrm{dBm}$. (e-f) The oscillation period
and the frequency shift versus pump power from $21.85\:\mathrm{dBm}$
to $23.1\:\mathrm{dBm}$. The solid line is the theoretical curve.}
\end{figure}

As the stabilization and control of the magnonic frequency comb are
critical for the potential application of high-precision magnonic
frequency spectroscopy, here, we also demonstrate the stabilization
and control of the frequency spacing of the magnonic frequency comb
via injection locking \cite{wan2020frequency}, which significantly
suppresses the instability of the comb teeth, as shown in Fig. 4(a).
An additional source is injected into the YIG microsphere at a frequency
of $4.71492\:\mathrm{GHz}$. Prior to turning on the external source,
the magnonic frequency comb is generated, and the comb tooth on the
RF spectrum is unstable (stage I), particularly when focusing on the
comb line at $4.71492\:\mathrm{GHz}$. We then turn on the external
source, and the stabilization of the magnonic frequency comb is notably
improved (stage II). Finally, by turning off the external source,
the stable comb tooth immediately returns to the initial state and
becomes as unstable as before (stage III). Furthermore, we study the
locking range of this injection locking scheme. Figure 4(b) displays
the evolution of the RF spectra with a power of $-6.4\:\mathrm{dBm}$
when the external source frequency slowly varied. Initially, when
the frequency difference between the external source and comb line
is relatively large, beat notes of the initial frequency and external
source frequency and their harmonic components all existe in the RF
spectrum. As we continue scanning, the comb is injection locked by
the external source when their frequency difference is very close.
Scanning in the same direction further led to the external source
frequency crossing the comb line and going out of the locking range
at last. The locking range increases with the enhancement of the external
source power, up to a maximum of $1.2\,\mathrm{kHz}$, as depicted
in Fig. 4(c).
\begin{figure}
\begin{centering}
\includegraphics[clip,width=1\columnwidth]{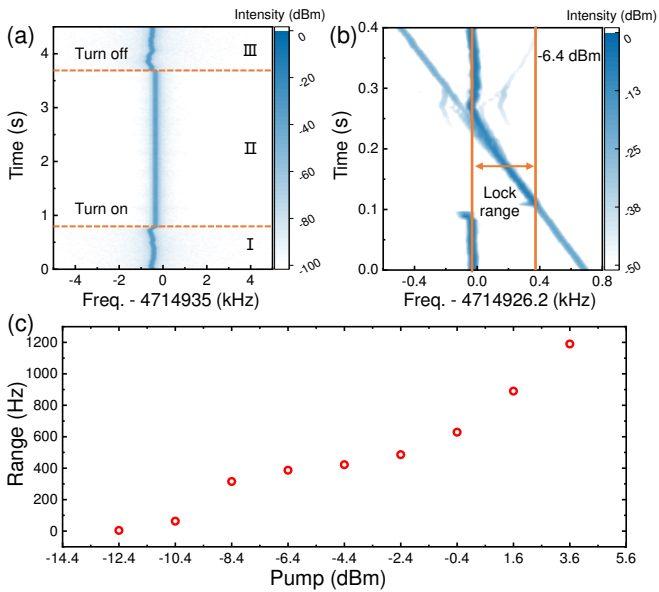}
\par\end{centering}
\caption{\label{fig:Fig4} (a) Evolution of one comb tooth when the external
source is on and off. The initial comb tooth is unstable (\mbox{I}).
The comb tooth is synchronized to the external source (\mbox{II}).
The comb tooth returns to the initial state after turning off the
external source (\mbox{III}). (b) Evolution of the RF spectrum with
varied injected frequency. The injected power is $-6.4\:\mathrm{dBm}$,
and the comb is synchronized to the external source when the frequency
difference is less than $400\ \mathrm{Hz}$. (c) Locking ranges with
varied pump power.}
\end{figure}

\subsection{Discussion}

In conclusion, we demonstrate a magnonic frequency comb in the magnomechanical
resonator. The magnonic frequency comb has up to 20 comb lines and
a frequency spacing $10.08\ \mathrm{MHz}$, equaling the resonant
frequency of the mechanical mode. Additionally, we study the thermal
effect involved in the frequency comb generation process, where the
frequency comb exhibits periodic oscillation dependent on pump power.
Our experimental results are in agreement with numerical calculations.
Furthermore, we demonstrate the stabilization and control of the frequency
spacing of the magnonic frequency comb via injection locking. We can
achieve tuning of comb teeth beyond a range of $1\ \mathrm{kHz}$.
Our work not only advances the study of nonlinear physics in magnonic
system but also unlocks the potential of magnonic frequency combs
for sensing and metrology.

\section*{Acknowledgments}
The authors thank C.-L. Zou for helpful discussions and suggestion.
The work was supported by the National Key Research and Development Program (Grant No. 2020YFB2205801) and National Natural Science Foundation
of China (Grant No. 12293052, 12293050, 11934012, 92050109, 12104442, and
92250302), Innovation program for Quantum Science and Technology
(2021ZD0303203), the CAS Project for Young Scientists in Basic Research
(YSBR-069), the Fundamental Research Funds for the Central Universities.
This work was partially carried out at the USTC Center for Micro and
Nanoscale Research and Fabrication.


\begin{thebibliography}{39}%
\makeatletter
\providecommand \@ifxundefined [1]{%
 \@ifx{#1\undefined}
}%
\providecommand \@ifnum [1]{%
 \ifnum #1\expandafter \@firstoftwo
 \else \expandafter \@secondoftwo
 \fi
}%
\providecommand \@ifx [1]{%
 \ifx #1\expandafter \@firstoftwo
 \else \expandafter \@secondoftwo
 \fi
}%
\providecommand \natexlab [1]{#1}%
\providecommand \enquote  [1]{``#1''}%
\providecommand \bibnamefont  [1]{#1}%
\providecommand \bibfnamefont [1]{#1}%
\providecommand \citenamefont [1]{#1}%
\providecommand \href@noop [0]{\@secondoftwo}%
\providecommand \href [0]{\begingroup \@sanitize@url \@href}%
\providecommand \@href[1]{\@@startlink{#1}\@@href}%
\providecommand \@@href[1]{\endgroup#1\@@endlink}%
\providecommand \@sanitize@url [0]{\catcode `\\12\catcode `\$12\catcode
  `\&12\catcode `\#12\catcode `\^12\catcode `\_12\catcode `\%12\relax}%
\providecommand \@@startlink[1]{}%
\providecommand \@@endlink[0]{}%
\providecommand \url  [0]{\begingroup\@sanitize@url \@url }%
\providecommand \@url [1]{\endgroup\@href {#1}{\urlprefix }}%
\providecommand \urlprefix  [0]{URL }%
\providecommand \Eprint [0]{\href }%
\providecommand \doibase [0]{http://dx.doi.org/}%
\providecommand \selectlanguage [0]{\@gobble}%
\providecommand \bibinfo  [0]{\@secondoftwo}%
\providecommand \bibfield  [0]{\@secondoftwo}%
\providecommand \translation [1]{[#1]}%
\providecommand \BibitemOpen [0]{}%
\providecommand \bibitemStop [0]{}%
\providecommand \bibitemNoStop [0]{.\EOS\space}%
\providecommand \EOS [0]{\spacefactor3000\relax}%
\providecommand \BibitemShut  [1]{\csname bibitem#1\endcsname}%
\let\auto@bib@innerbib\@empty
\bibitem [{\citenamefont {Kippenberg}\ \emph {et~al.}(2011)\citenamefont
  {Kippenberg}, \citenamefont {Holzwarth},\ and\ \citenamefont
  {Diddams}}]{kippenberg2011microresonator}%
  \BibitemOpen
  \bibfield  {author} {\bibinfo {author} {\bibfnamefont {T.~J.}\ \bibnamefont
  {Kippenberg}}, \bibinfo {author} {\bibfnamefont {R.}~\bibnamefont
  {Holzwarth}}, \ and\ \bibinfo {author} {\bibfnamefont {S.~A.}\ \bibnamefont
  {Diddams}},\ }\bibfield  {title} {\enquote {\bibinfo {title}
  {Microresonator-based optical frequency combs},}\ }\href@noop {} {\bibfield
  {journal} {\bibinfo  {journal} {Science}\ }\textbf {\bibinfo {volume}
  {332}},\ \bibinfo {pages} {555} (\bibinfo {year} {2011})}\BibitemShut
  {NoStop}%
\bibitem [{\citenamefont {Kippenberg}\ \emph {et~al.}(2018)\citenamefont
  {Kippenberg}, \citenamefont {Gaeta}, \citenamefont {Lipson},\ and\
  \citenamefont {Gorodetsky}}]{kippenberg2018dissipative}%
  \BibitemOpen
  \bibfield  {author} {\bibinfo {author} {\bibfnamefont {T.~J.}\ \bibnamefont
  {Kippenberg}}, \bibinfo {author} {\bibfnamefont {A.~L.}\ \bibnamefont
  {Gaeta}}, \bibinfo {author} {\bibfnamefont {M.}~\bibnamefont {Lipson}}, \
  and\ \bibinfo {author} {\bibfnamefont {M.~L.}\ \bibnamefont {Gorodetsky}},\
  }\bibfield  {title} {\enquote {\bibinfo {title} {Dissipative kerr solitons in
  optical microresonators},}\ }\href@noop {} {\bibfield  {journal} {\bibinfo
  {journal} {Science}\ }\textbf {\bibinfo {volume} {361}},\ \bibinfo {pages}
  {eaan8083} (\bibinfo {year} {2018})}\BibitemShut {NoStop}%
\bibitem [{\citenamefont {Diddams}\ \emph {et~al.}(2020)\citenamefont
  {Diddams}, \citenamefont {Vahala},\ and\ \citenamefont
  {Udem}}]{diddams2020optical}%
  \BibitemOpen
  \bibfield  {author} {\bibinfo {author} {\bibfnamefont {S.~A.}\ \bibnamefont
  {Diddams}}, \bibinfo {author} {\bibfnamefont {K.}~\bibnamefont {Vahala}}, \
  and\ \bibinfo {author} {\bibfnamefont {T.}~\bibnamefont {Udem}},\ }\bibfield
  {title} {\enquote {\bibinfo {title} {Optical frequency combs: Coherently
  uniting the electromagnetic spectrum},}\ }\href@noop {} {\bibfield  {journal}
  {\bibinfo  {journal} {Science}\ }\textbf {\bibinfo {volume} {369}},\ \bibinfo
  {pages} {eaay3676} (\bibinfo {year} {2020})}\BibitemShut {NoStop}%
\bibitem [{\citenamefont {Newman}\ \emph {et~al.}(2019)\citenamefont {Newman},
  \citenamefont {Maurice}, \citenamefont {Drake}, \citenamefont {Stone},
  \citenamefont {Briles}, \citenamefont {Spencer}, \citenamefont {Fredrick},
  \citenamefont {Li}, \citenamefont {Westly}, \citenamefont {Ilic} \emph
  {et~al.}}]{newman2019architecture}%
  \BibitemOpen
  \bibfield  {author} {\bibinfo {author} {\bibfnamefont {Z.~L.}\ \bibnamefont
  {Newman}}, \bibinfo {author} {\bibfnamefont {V.}~\bibnamefont {Maurice}},
  \bibinfo {author} {\bibfnamefont {T.}~\bibnamefont {Drake}}, \bibinfo
  {author} {\bibfnamefont {J.~R.}\ \bibnamefont {Stone}}, \bibinfo {author}
  {\bibfnamefont {T.~C.}\ \bibnamefont {Briles}}, \bibinfo {author}
  {\bibfnamefont {D.~T.}\ \bibnamefont {Spencer}}, \bibinfo {author}
  {\bibfnamefont {C.}~\bibnamefont {Fredrick}}, \bibinfo {author}
  {\bibfnamefont {Q.}~\bibnamefont {Li}}, \bibinfo {author} {\bibfnamefont
  {D.}~\bibnamefont {Westly}}, \bibinfo {author} {\bibfnamefont {B.~R.}\
  \bibnamefont {Ilic}},  \emph {et~al.},\ }\bibfield  {title} {\enquote
  {\bibinfo {title} {Architecture for the photonic integration of an optical
  atomic clock},}\ }\href@noop {} {\bibfield  {journal} {\bibinfo  {journal}
  {Optica}\ }\textbf {\bibinfo {volume} {6}},\ \bibinfo {pages} {680} (\bibinfo
  {year} {2019})}\BibitemShut {NoStop}%
\bibitem [{\citenamefont {Suh}\ and\ \citenamefont
  {Vahala}(2018)}]{suh2018soliton}%
  \BibitemOpen
  \bibfield  {author} {\bibinfo {author} {\bibfnamefont {M.-G.}\ \bibnamefont
  {Suh}}\ and\ \bibinfo {author} {\bibfnamefont {K.~J.}\ \bibnamefont
  {Vahala}},\ }\bibfield  {title} {\enquote {\bibinfo {title} {Soliton
  microcomb range measurement},}\ }\href@noop {} {\bibfield  {journal}
  {\bibinfo  {journal} {Science}\ }\textbf {\bibinfo {volume} {359}},\ \bibinfo
  {pages} {884} (\bibinfo {year} {2018})}\BibitemShut {NoStop}%
\bibitem [{\citenamefont {Trocha}\ \emph {et~al.}(2018)\citenamefont {Trocha},
  \citenamefont {Karpov}, \citenamefont {Ganin}, \citenamefont {Pfeiffer},
  \citenamefont {Kordts}, \citenamefont {Wolf}, \citenamefont {Krockenberger},
  \citenamefont {Marin-Palomo}, \citenamefont {Weimann}, \citenamefont {Randel}
  \emph {et~al.}}]{trocha2018ultrafast}%
  \BibitemOpen
  \bibfield  {author} {\bibinfo {author} {\bibfnamefont {P.}~\bibnamefont
  {Trocha}}, \bibinfo {author} {\bibfnamefont {M.}~\bibnamefont {Karpov}},
  \bibinfo {author} {\bibfnamefont {D.}~\bibnamefont {Ganin}}, \bibinfo
  {author} {\bibfnamefont {M.~H.}\ \bibnamefont {Pfeiffer}}, \bibinfo {author}
  {\bibfnamefont {A.}~\bibnamefont {Kordts}}, \bibinfo {author} {\bibfnamefont
  {S.}~\bibnamefont {Wolf}}, \bibinfo {author} {\bibfnamefont {J.}~\bibnamefont
  {Krockenberger}}, \bibinfo {author} {\bibfnamefont {P.}~\bibnamefont
  {Marin-Palomo}}, \bibinfo {author} {\bibfnamefont {C.}~\bibnamefont
  {Weimann}}, \bibinfo {author} {\bibfnamefont {S.}~\bibnamefont {Randel}},
  \emph {et~al.},\ }\bibfield  {title} {\enquote {\bibinfo {title} {Ultrafast
  optical ranging using microresonator soliton frequency combs},}\ }\href@noop
  {} {\bibfield  {journal} {\bibinfo  {journal} {Science}\ }\textbf {\bibinfo
  {volume} {359}},\ \bibinfo {pages} {887} (\bibinfo {year}
  {2018})}\BibitemShut {NoStop}%
\bibitem [{\citenamefont {Wang}\ \emph
  {et~al.}(2020{\natexlab{a}})\citenamefont {Wang}, \citenamefont {Lu},
  \citenamefont {Wang}, \citenamefont {Zhang}, \citenamefont {Chen},
  \citenamefont {Wang}, \citenamefont {Zheng}, \citenamefont {Chu},
  \citenamefont {Zhao}, \citenamefont {Little} \emph {et~al.}}]{wang2020long}%
  \BibitemOpen
  \bibfield  {author} {\bibinfo {author} {\bibfnamefont {J.}~\bibnamefont
  {Wang}}, \bibinfo {author} {\bibfnamefont {Z.}~\bibnamefont {Lu}}, \bibinfo
  {author} {\bibfnamefont {W.}~\bibnamefont {Wang}}, \bibinfo {author}
  {\bibfnamefont {F.}~\bibnamefont {Zhang}}, \bibinfo {author} {\bibfnamefont
  {J.}~\bibnamefont {Chen}}, \bibinfo {author} {\bibfnamefont {Y.}~\bibnamefont
  {Wang}}, \bibinfo {author} {\bibfnamefont {J.}~\bibnamefont {Zheng}},
  \bibinfo {author} {\bibfnamefont {S.~T.}\ \bibnamefont {Chu}}, \bibinfo
  {author} {\bibfnamefont {W.}~\bibnamefont {Zhao}}, \bibinfo {author}
  {\bibfnamefont {B.~E.}\ \bibnamefont {Little}},  \emph {et~al.},\ }\bibfield
  {title} {\enquote {\bibinfo {title} {Long-distance ranging with high
  precision using a soliton microcomb},}\ }\href@noop {} {\bibfield  {journal}
  {\bibinfo  {journal} {Photonics Research}\ }\textbf {\bibinfo {volume} {8}},\
  \bibinfo {pages} {1964} (\bibinfo {year} {2020}{\natexlab{a}})}\BibitemShut
  {NoStop}%
\bibitem [{\citenamefont {Liang}\ \emph {et~al.}(2015)\citenamefont {Liang},
  \citenamefont {Eliyahu}, \citenamefont {Ilchenko}, \citenamefont
  {Savchenkov}, \citenamefont {Matsko}, \citenamefont {Seidel},\ and\
  \citenamefont {Maleki}}]{liang2015high}%
  \BibitemOpen
  \bibfield  {author} {\bibinfo {author} {\bibfnamefont {W.}~\bibnamefont
  {Liang}}, \bibinfo {author} {\bibfnamefont {D.}~\bibnamefont {Eliyahu}},
  \bibinfo {author} {\bibfnamefont {V.~S.}\ \bibnamefont {Ilchenko}}, \bibinfo
  {author} {\bibfnamefont {A.~A.}\ \bibnamefont {Savchenkov}}, \bibinfo
  {author} {\bibfnamefont {A.~B.}\ \bibnamefont {Matsko}}, \bibinfo {author}
  {\bibfnamefont {D.}~\bibnamefont {Seidel}}, \ and\ \bibinfo {author}
  {\bibfnamefont {L.}~\bibnamefont {Maleki}},\ }\bibfield  {title} {\enquote
  {\bibinfo {title} {High spectral purity kerr frequency comb radio frequency
  photonic oscillator},}\ }\href@noop {} {\bibfield  {journal} {\bibinfo
  {journal} {Nature Communications}\ }\textbf {\bibinfo {volume} {6}},\
  \bibinfo {pages} {7957} (\bibinfo {year} {2015})}\BibitemShut {NoStop}%
\bibitem [{\citenamefont {Lucas}\ \emph {et~al.}(2020)\citenamefont {Lucas},
  \citenamefont {Brochard}, \citenamefont {Bouchand}, \citenamefont {Schilt},
  \citenamefont {S{\"u}dmeyer},\ and\ \citenamefont
  {Kippenberg}}]{lucas2020ultralow}%
  \BibitemOpen
  \bibfield  {author} {\bibinfo {author} {\bibfnamefont {E.}~\bibnamefont
  {Lucas}}, \bibinfo {author} {\bibfnamefont {P.}~\bibnamefont {Brochard}},
  \bibinfo {author} {\bibfnamefont {R.}~\bibnamefont {Bouchand}}, \bibinfo
  {author} {\bibfnamefont {S.}~\bibnamefont {Schilt}}, \bibinfo {author}
  {\bibfnamefont {T.}~\bibnamefont {S{\"u}dmeyer}}, \ and\ \bibinfo {author}
  {\bibfnamefont {T.~J.}\ \bibnamefont {Kippenberg}},\ }\bibfield  {title}
  {\enquote {\bibinfo {title} {Ultralow-noise photonic microwave synthesis
  using a soliton microcomb-based transfer oscillator},}\ }\href@noop {}
  {\bibfield  {journal} {\bibinfo  {journal} {Nature Communications}\ }\textbf
  {\bibinfo {volume} {11}},\ \bibinfo {pages} {374} (\bibinfo {year}
  {2020})}\BibitemShut {NoStop}%
\bibitem [{\citenamefont {Liu}\ \emph {et~al.}(2020)\citenamefont {Liu},
  \citenamefont {Lucas}, \citenamefont {Raja}, \citenamefont {He},
  \citenamefont {Riemensberger}, \citenamefont {Wang}, \citenamefont {Karpov},
  \citenamefont {Guo}, \citenamefont {Bouchand},\ and\ \citenamefont
  {Kippenberg}}]{liu2020photonic}%
  \BibitemOpen
  \bibfield  {author} {\bibinfo {author} {\bibfnamefont {J.}~\bibnamefont
  {Liu}}, \bibinfo {author} {\bibfnamefont {E.}~\bibnamefont {Lucas}}, \bibinfo
  {author} {\bibfnamefont {A.~S.}\ \bibnamefont {Raja}}, \bibinfo {author}
  {\bibfnamefont {J.}~\bibnamefont {He}}, \bibinfo {author} {\bibfnamefont
  {J.}~\bibnamefont {Riemensberger}}, \bibinfo {author} {\bibfnamefont {R.~N.}\
  \bibnamefont {Wang}}, \bibinfo {author} {\bibfnamefont {M.}~\bibnamefont
  {Karpov}}, \bibinfo {author} {\bibfnamefont {H.}~\bibnamefont {Guo}},
  \bibinfo {author} {\bibfnamefont {R.}~\bibnamefont {Bouchand}}, \ and\
  \bibinfo {author} {\bibfnamefont {T.~J.}\ \bibnamefont {Kippenberg}},\
  }\bibfield  {title} {\enquote {\bibinfo {title} {Photonic microwave
  generation in the x-and k-band using integrated soliton microcombs},}\
  }\href@noop {} {\bibfield  {journal} {\bibinfo  {journal} {Nature Photonics}\
  }\textbf {\bibinfo {volume} {14}},\ \bibinfo {pages} {486} (\bibinfo {year}
  {2020})}\BibitemShut {NoStop}%
\bibitem [{\citenamefont {Marin-Palomo}\ \emph {et~al.}(2017)\citenamefont
  {Marin-Palomo}, \citenamefont {Kemal}, \citenamefont {Karpov}, \citenamefont
  {Kordts}, \citenamefont {Pfeifle}, \citenamefont {Pfeiffer}, \citenamefont
  {Trocha}, \citenamefont {Wolf}, \citenamefont {Brasch}, \citenamefont
  {Anderson} \emph {et~al.}}]{marin2017microresonator}%
  \BibitemOpen
  \bibfield  {author} {\bibinfo {author} {\bibfnamefont {P.}~\bibnamefont
  {Marin-Palomo}}, \bibinfo {author} {\bibfnamefont {J.~N.}\ \bibnamefont
  {Kemal}}, \bibinfo {author} {\bibfnamefont {M.}~\bibnamefont {Karpov}},
  \bibinfo {author} {\bibfnamefont {A.}~\bibnamefont {Kordts}}, \bibinfo
  {author} {\bibfnamefont {J.}~\bibnamefont {Pfeifle}}, \bibinfo {author}
  {\bibfnamefont {M.~H.}\ \bibnamefont {Pfeiffer}}, \bibinfo {author}
  {\bibfnamefont {P.}~\bibnamefont {Trocha}}, \bibinfo {author} {\bibfnamefont
  {S.}~\bibnamefont {Wolf}}, \bibinfo {author} {\bibfnamefont {V.}~\bibnamefont
  {Brasch}}, \bibinfo {author} {\bibfnamefont {M.~H.}\ \bibnamefont
  {Anderson}},  \emph {et~al.},\ }\bibfield  {title} {\enquote {\bibinfo
  {title} {Microresonator-based solitons for massively parallel coherent
  optical communications},}\ }\href@noop {} {\bibfield  {journal} {\bibinfo
  {journal} {Nature}\ }\textbf {\bibinfo {volume} {546}},\ \bibinfo {pages}
  {274} (\bibinfo {year} {2017})}\BibitemShut {NoStop}%
\bibitem [{\citenamefont {Corcoran}\ \emph {et~al.}(2020)\citenamefont
  {Corcoran}, \citenamefont {Tan}, \citenamefont {Xu}, \citenamefont {Boes},
  \citenamefont {Wu}, \citenamefont {Nguyen}, \citenamefont {Chu},
  \citenamefont {Little}, \citenamefont {Morandotti}, \citenamefont {Mitchell}
  \emph {et~al.}}]{corcoran2020ultra}%
  \BibitemOpen
  \bibfield  {author} {\bibinfo {author} {\bibfnamefont {B.}~\bibnamefont
  {Corcoran}}, \bibinfo {author} {\bibfnamefont {M.}~\bibnamefont {Tan}},
  \bibinfo {author} {\bibfnamefont {X.}~\bibnamefont {Xu}}, \bibinfo {author}
  {\bibfnamefont {A.}~\bibnamefont {Boes}}, \bibinfo {author} {\bibfnamefont
  {J.}~\bibnamefont {Wu}}, \bibinfo {author} {\bibfnamefont {T.~G.}\
  \bibnamefont {Nguyen}}, \bibinfo {author} {\bibfnamefont {S.~T.}\
  \bibnamefont {Chu}}, \bibinfo {author} {\bibfnamefont {B.~E.}\ \bibnamefont
  {Little}}, \bibinfo {author} {\bibfnamefont {R.}~\bibnamefont {Morandotti}},
  \bibinfo {author} {\bibfnamefont {A.}~\bibnamefont {Mitchell}},  \emph
  {et~al.},\ }\bibfield  {title} {\enquote {\bibinfo {title} {Ultra-dense
  optical data transmission over standard fibre with a single chip source},}\
  }\href@noop {} {\bibfield  {journal} {\bibinfo  {journal} {Nature
  Communications}\ }\textbf {\bibinfo {volume} {11}},\ \bibinfo {pages} {2568}
  (\bibinfo {year} {2020})}\BibitemShut {NoStop}%
\bibitem [{\citenamefont {Wang}\ \emph
  {et~al.}(2020{\natexlab{b}})\citenamefont {Wang}, \citenamefont {Wang},
  \citenamefont {Niu}, \citenamefont {Wang}, \citenamefont {Zou}, \citenamefont
  {Dong}, \citenamefont {Little}, \citenamefont {Chu}, \citenamefont {Liu},
  \citenamefont {Hao} \emph {et~al.}}]{wang2020quantum}%
  \BibitemOpen
  \bibfield  {author} {\bibinfo {author} {\bibfnamefont {F.-X.}\ \bibnamefont
  {Wang}}, \bibinfo {author} {\bibfnamefont {W.}~\bibnamefont {Wang}}, \bibinfo
  {author} {\bibfnamefont {R.}~\bibnamefont {Niu}}, \bibinfo {author}
  {\bibfnamefont {X.}~\bibnamefont {Wang}}, \bibinfo {author} {\bibfnamefont
  {C.-L.}\ \bibnamefont {Zou}}, \bibinfo {author} {\bibfnamefont {C.-H.}\
  \bibnamefont {Dong}}, \bibinfo {author} {\bibfnamefont {B.~E.}\ \bibnamefont
  {Little}}, \bibinfo {author} {\bibfnamefont {S.~T.}\ \bibnamefont {Chu}},
  \bibinfo {author} {\bibfnamefont {H.}~\bibnamefont {Liu}}, \bibinfo {author}
  {\bibfnamefont {P.}~\bibnamefont {Hao}},  \emph {et~al.},\ }\bibfield
  {title} {\enquote {\bibinfo {title} {Quantum key distribution with on-chip
  dissipative kerr soliton},}\ }\href@noop {} {\bibfield  {journal} {\bibinfo
  {journal} {Laser \& Photonics Reviews}\ }\textbf {\bibinfo {volume} {14}},\
  \bibinfo {pages} {1900190} (\bibinfo {year}
  {2020}{\natexlab{b}})}\BibitemShut {NoStop}%
\bibitem [{\citenamefont {Suh}\ \emph {et~al.}(2016)\citenamefont {Suh},
  \citenamefont {Yang}, \citenamefont {Yang}, \citenamefont {Yi},\ and\
  \citenamefont {Vahala}}]{suh2016microresonator}%
  \BibitemOpen
  \bibfield  {author} {\bibinfo {author} {\bibfnamefont {M.-G.}\ \bibnamefont
  {Suh}}, \bibinfo {author} {\bibfnamefont {Q.-F.}\ \bibnamefont {Yang}},
  \bibinfo {author} {\bibfnamefont {K.~Y.}\ \bibnamefont {Yang}}, \bibinfo
  {author} {\bibfnamefont {X.}~\bibnamefont {Yi}}, \ and\ \bibinfo {author}
  {\bibfnamefont {K.~J.}\ \bibnamefont {Vahala}},\ }\bibfield  {title}
  {\enquote {\bibinfo {title} {Microresonator soliton dual-comb
  spectroscopy},}\ }\href@noop {} {\bibfield  {journal} {\bibinfo  {journal}
  {Science}\ }\textbf {\bibinfo {volume} {354}},\ \bibinfo {pages} {600}
  (\bibinfo {year} {2016})}\BibitemShut {NoStop}%
\bibitem [{\citenamefont {Yu}\ \emph {et~al.}(2018)\citenamefont {Yu},
  \citenamefont {Okawachi}, \citenamefont {Griffith}, \citenamefont
  {Picqu{\'e}}, \citenamefont {Lipson},\ and\ \citenamefont
  {Gaeta}}]{yu2018silicon}%
  \BibitemOpen
  \bibfield  {author} {\bibinfo {author} {\bibfnamefont {M.}~\bibnamefont
  {Yu}}, \bibinfo {author} {\bibfnamefont {Y.}~\bibnamefont {Okawachi}},
  \bibinfo {author} {\bibfnamefont {A.~G.}\ \bibnamefont {Griffith}}, \bibinfo
  {author} {\bibfnamefont {N.}~\bibnamefont {Picqu{\'e}}}, \bibinfo {author}
  {\bibfnamefont {M.}~\bibnamefont {Lipson}}, \ and\ \bibinfo {author}
  {\bibfnamefont {A.~L.}\ \bibnamefont {Gaeta}},\ }\bibfield  {title} {\enquote
  {\bibinfo {title} {Silicon-chip-based mid-infrared dual-comb spectroscopy},}\
  }\href@noop {} {\bibfield  {journal} {\bibinfo  {journal} {Nature
  Communications}\ }\textbf {\bibinfo {volume} {9}},\ \bibinfo {pages} {1869}
  (\bibinfo {year} {2018})}\BibitemShut {NoStop}%
\bibitem [{\citenamefont {Yang}\ \emph {et~al.}(2019)\citenamefont {Yang},
  \citenamefont {Shen}, \citenamefont {Wang}, \citenamefont {Tran},
  \citenamefont {Zhang}, \citenamefont {Yang}, \citenamefont {Wu},
  \citenamefont {Bao}, \citenamefont {Bowers}, \citenamefont {Yariv} \emph
  {et~al.}}]{yang2019vernier}%
  \BibitemOpen
  \bibfield  {author} {\bibinfo {author} {\bibfnamefont {Q.-F.}\ \bibnamefont
  {Yang}}, \bibinfo {author} {\bibfnamefont {B.}~\bibnamefont {Shen}}, \bibinfo
  {author} {\bibfnamefont {H.}~\bibnamefont {Wang}}, \bibinfo {author}
  {\bibfnamefont {M.}~\bibnamefont {Tran}}, \bibinfo {author} {\bibfnamefont
  {Z.}~\bibnamefont {Zhang}}, \bibinfo {author} {\bibfnamefont {K.~Y.}\
  \bibnamefont {Yang}}, \bibinfo {author} {\bibfnamefont {L.}~\bibnamefont
  {Wu}}, \bibinfo {author} {\bibfnamefont {C.}~\bibnamefont {Bao}}, \bibinfo
  {author} {\bibfnamefont {J.}~\bibnamefont {Bowers}}, \bibinfo {author}
  {\bibfnamefont {A.}~\bibnamefont {Yariv}},  \emph {et~al.},\ }\bibfield
  {title} {\enquote {\bibinfo {title} {Vernier spectrometer using
  counterpropagating soliton microcombs},}\ }\href@noop {} {\bibfield
  {journal} {\bibinfo  {journal} {Science}\ }\textbf {\bibinfo {volume}
  {363}},\ \bibinfo {pages} {965} (\bibinfo {year} {2019})}\BibitemShut
  {NoStop}%
\bibitem [{\citenamefont {Niu}\ \emph {et~al.}(2023)\citenamefont {Niu},
  \citenamefont {Li}, \citenamefont {Wan}, \citenamefont {Sun}, \citenamefont
  {Hu}, \citenamefont {Zou}, \citenamefont {Guo},\ and\ \citenamefont
  {Dong}}]{niu2023khz}%
  \BibitemOpen
  \bibfield  {author} {\bibinfo {author} {\bibfnamefont {R.}~\bibnamefont
  {Niu}}, \bibinfo {author} {\bibfnamefont {M.}~\bibnamefont {Li}}, \bibinfo
  {author} {\bibfnamefont {S.}~\bibnamefont {Wan}}, \bibinfo {author}
  {\bibfnamefont {Y.~R.}\ \bibnamefont {Sun}}, \bibinfo {author} {\bibfnamefont
  {S.-M.}\ \bibnamefont {Hu}}, \bibinfo {author} {\bibfnamefont {C.-L.}\
  \bibnamefont {Zou}}, \bibinfo {author} {\bibfnamefont {G.-C.}\ \bibnamefont
  {Guo}}, \ and\ \bibinfo {author} {\bibfnamefont {C.-H.}\ \bibnamefont
  {Dong}},\ }\bibfield  {title} {\enquote {\bibinfo {title} {khz-precision
  wavemeter based on reconfigurable microsoliton},}\ }\href@noop {} {\bibfield
  {journal} {\bibinfo  {journal} {Nature Communications}\ }\textbf {\bibinfo
  {volume} {14}},\ \bibinfo {pages} {169} (\bibinfo {year} {2023})}\BibitemShut
  {NoStop}%
\bibitem [{\citenamefont {Ji}\ \emph {et~al.}(2019)\citenamefont {Ji},
  \citenamefont {Yao}, \citenamefont {Klenner}, \citenamefont {Gan},
  \citenamefont {Gaeta}, \citenamefont {Hendon},\ and\ \citenamefont
  {Lipson}}]{ji2019chip}%
  \BibitemOpen
  \bibfield  {author} {\bibinfo {author} {\bibfnamefont {X.}~\bibnamefont
  {Ji}}, \bibinfo {author} {\bibfnamefont {X.}~\bibnamefont {Yao}}, \bibinfo
  {author} {\bibfnamefont {A.}~\bibnamefont {Klenner}}, \bibinfo {author}
  {\bibfnamefont {Y.}~\bibnamefont {Gan}}, \bibinfo {author} {\bibfnamefont
  {A.~L.}\ \bibnamefont {Gaeta}}, \bibinfo {author} {\bibfnamefont {C.~P.}\
  \bibnamefont {Hendon}}, \ and\ \bibinfo {author} {\bibfnamefont
  {M.}~\bibnamefont {Lipson}},\ }\bibfield  {title} {\enquote {\bibinfo {title}
  {Chip-based frequency comb sources for optical coherence tomography},}\
  }\href@noop {} {\bibfield  {journal} {\bibinfo  {journal} {Optics Express}\
  }\textbf {\bibinfo {volume} {27}},\ \bibinfo {pages} {19896} (\bibinfo {year}
  {2019})}\BibitemShut {NoStop}%
\bibitem [{\citenamefont {Marchand}\ \emph {et~al.}(2021)\citenamefont
  {Marchand}, \citenamefont {Riemensberger}, \citenamefont {Skehan},
  \citenamefont {Ho}, \citenamefont {Pfeiffer}, \citenamefont {Liu},
  \citenamefont {Hauger}, \citenamefont {Lasser},\ and\ \citenamefont
  {Kippenberg}}]{marchand2021soliton}%
  \BibitemOpen
  \bibfield  {author} {\bibinfo {author} {\bibfnamefont {P.~J.}\ \bibnamefont
  {Marchand}}, \bibinfo {author} {\bibfnamefont {J.}~\bibnamefont
  {Riemensberger}}, \bibinfo {author} {\bibfnamefont {J.~C.}\ \bibnamefont
  {Skehan}}, \bibinfo {author} {\bibfnamefont {J.-J.}\ \bibnamefont {Ho}},
  \bibinfo {author} {\bibfnamefont {M.~H.}\ \bibnamefont {Pfeiffer}}, \bibinfo
  {author} {\bibfnamefont {J.}~\bibnamefont {Liu}}, \bibinfo {author}
  {\bibfnamefont {C.}~\bibnamefont {Hauger}}, \bibinfo {author} {\bibfnamefont
  {T.}~\bibnamefont {Lasser}}, \ and\ \bibinfo {author} {\bibfnamefont {T.~J.}\
  \bibnamefont {Kippenberg}},\ }\bibfield  {title} {\enquote {\bibinfo {title}
  {Soliton microcomb based spectral domain optical coherence tomography},}\
  }\href@noop {} {\bibfield  {journal} {\bibinfo  {journal} {Nature
  Communications}\ }\textbf {\bibinfo {volume} {12}},\ \bibinfo {pages} {427}
  (\bibinfo {year} {2021})}\BibitemShut {NoStop}%
\bibitem [{\citenamefont {Cao}\ \emph {et~al.}(2014)\citenamefont {Cao},
  \citenamefont {Qi}, \citenamefont {Peng}, \citenamefont {Wang},\ and\
  \citenamefont {Schmelcher}}]{cao2014phononic}%
  \BibitemOpen
  \bibfield  {author} {\bibinfo {author} {\bibfnamefont {L.}~\bibnamefont
  {Cao}}, \bibinfo {author} {\bibfnamefont {D.}~\bibnamefont {Qi}}, \bibinfo
  {author} {\bibfnamefont {R.}~\bibnamefont {Peng}}, \bibinfo {author}
  {\bibfnamefont {M.}~\bibnamefont {Wang}}, \ and\ \bibinfo {author}
  {\bibfnamefont {P.}~\bibnamefont {Schmelcher}},\ }\bibfield  {title}
  {\enquote {\bibinfo {title} {Phononic frequency combs through nonlinear
  resonances},}\ }\href@noop {} {\bibfield  {journal} {\bibinfo  {journal}
  {Physical Review Letters}\ }\textbf {\bibinfo {volume} {112}},\ \bibinfo
  {pages} {075505} (\bibinfo {year} {2014})}\BibitemShut {NoStop}%
\bibitem [{\citenamefont {de~Jong}\ \emph {et~al.}(2023)\citenamefont
  {de~Jong}, \citenamefont {Ganesan}, \citenamefont {Cupertino}, \citenamefont
  {Gr{\"o}blacher},\ and\ \citenamefont {Norte}}]{de2023mechanical}%
  \BibitemOpen
  \bibfield  {author} {\bibinfo {author} {\bibfnamefont {M.~H.}\ \bibnamefont
  {de~Jong}}, \bibinfo {author} {\bibfnamefont {A.}~\bibnamefont {Ganesan}},
  \bibinfo {author} {\bibfnamefont {A.}~\bibnamefont {Cupertino}}, \bibinfo
  {author} {\bibfnamefont {S.}~\bibnamefont {Gr{\"o}blacher}}, \ and\ \bibinfo
  {author} {\bibfnamefont {R.~A.}\ \bibnamefont {Norte}},\ }\bibfield  {title}
  {\enquote {\bibinfo {title} {Mechanical overtone frequency combs},}\
  }\href@noop {} {\bibfield  {journal} {\bibinfo  {journal} {Nature
  Communications}\ }\textbf {\bibinfo {volume} {14}},\ \bibinfo {pages} {1458}
  (\bibinfo {year} {2023})}\BibitemShut {NoStop}%
\bibitem [{\citenamefont {Ganesan}\ \emph {et~al.}(2017)\citenamefont
  {Ganesan}, \citenamefont {Do},\ and\ \citenamefont
  {Seshia}}]{ganesan2017phononic}%
  \BibitemOpen
  \bibfield  {author} {\bibinfo {author} {\bibfnamefont {A.}~\bibnamefont
  {Ganesan}}, \bibinfo {author} {\bibfnamefont {C.}~\bibnamefont {Do}}, \ and\
  \bibinfo {author} {\bibfnamefont {A.}~\bibnamefont {Seshia}},\ }\bibfield
  {title} {\enquote {\bibinfo {title} {Phononic frequency comb via intrinsic
  three-wave mixing},}\ }\href@noop {} {\bibfield  {journal} {\bibinfo
  {journal} {Physical Review Letters}\ }\textbf {\bibinfo {volume} {118}},\
  \bibinfo {pages} {033903} (\bibinfo {year} {2017})}\BibitemShut {NoStop}%
\bibitem [{\citenamefont {Miri}\ \emph {et~al.}(2018)\citenamefont {Miri},
  \citenamefont {D’Aguanno},\ and\ \citenamefont
  {Al{\`u}}}]{miri2018optomechanical}%
  \BibitemOpen
  \bibfield  {author} {\bibinfo {author} {\bibfnamefont {M.-A.}\ \bibnamefont
  {Miri}}, \bibinfo {author} {\bibfnamefont {G.}~\bibnamefont {D’Aguanno}}, \
  and\ \bibinfo {author} {\bibfnamefont {A.}~\bibnamefont {Al{\`u}}},\
  }\bibfield  {title} {\enquote {\bibinfo {title} {Optomechanical frequency
  combs},}\ }\href@noop {} {\bibfield  {journal} {\bibinfo  {journal} {New
  Journal of Physics}\ }\textbf {\bibinfo {volume} {20}},\ \bibinfo {pages}
  {043013} (\bibinfo {year} {2018})}\BibitemShut {NoStop}%
\bibitem [{\citenamefont {Hu}\ \emph {et~al.}(2021)\citenamefont {Hu},
  \citenamefont {Ding}, \citenamefont {Qin}, \citenamefont {Gu}, \citenamefont
  {Wan}, \citenamefont {Xiao},\ and\ \citenamefont {Jiang}}]{hu2021generation}%
  \BibitemOpen
  \bibfield  {author} {\bibinfo {author} {\bibfnamefont {Y.}~\bibnamefont
  {Hu}}, \bibinfo {author} {\bibfnamefont {S.}~\bibnamefont {Ding}}, \bibinfo
  {author} {\bibfnamefont {Y.}~\bibnamefont {Qin}}, \bibinfo {author}
  {\bibfnamefont {J.}~\bibnamefont {Gu}}, \bibinfo {author} {\bibfnamefont
  {W.}~\bibnamefont {Wan}}, \bibinfo {author} {\bibfnamefont {M.}~\bibnamefont
  {Xiao}}, \ and\ \bibinfo {author} {\bibfnamefont {X.}~\bibnamefont {Jiang}},\
  }\bibfield  {title} {\enquote {\bibinfo {title} {Generation of optical
  frequency comb via giant optomechanical oscillation},}\ }\href@noop {}
  {\bibfield  {journal} {\bibinfo  {journal} {Physical Review Letters}\
  }\textbf {\bibinfo {volume} {127}},\ \bibinfo {pages} {134301} (\bibinfo
  {year} {2021})}\BibitemShut {NoStop}%
\bibitem [{\citenamefont {Zhang}\ \emph {et~al.}(2021)\citenamefont {Zhang},
  \citenamefont {Peng}, \citenamefont {Kim}, \citenamefont {Monifi},
  \citenamefont {Jiang}, \citenamefont {Li}, \citenamefont {Yu}, \citenamefont
  {Liu}, \citenamefont {Liu}, \citenamefont {Al{\`u}} \emph
  {et~al.}}]{zhang2021optomechanical}%
  \BibitemOpen
  \bibfield  {author} {\bibinfo {author} {\bibfnamefont {J.}~\bibnamefont
  {Zhang}}, \bibinfo {author} {\bibfnamefont {B.}~\bibnamefont {Peng}},
  \bibinfo {author} {\bibfnamefont {S.}~\bibnamefont {Kim}}, \bibinfo {author}
  {\bibfnamefont {F.}~\bibnamefont {Monifi}}, \bibinfo {author} {\bibfnamefont
  {X.}~\bibnamefont {Jiang}}, \bibinfo {author} {\bibfnamefont
  {Y.}~\bibnamefont {Li}}, \bibinfo {author} {\bibfnamefont {P.}~\bibnamefont
  {Yu}}, \bibinfo {author} {\bibfnamefont {L.}~\bibnamefont {Liu}}, \bibinfo
  {author} {\bibfnamefont {Y.-x.}\ \bibnamefont {Liu}}, \bibinfo {author}
  {\bibfnamefont {A.}~\bibnamefont {Al{\`u}}},  \emph {et~al.},\ }\bibfield
  {title} {\enquote {\bibinfo {title} {Optomechanical dissipative solitons},}\
  }\href@noop {} {\bibfield  {journal} {\bibinfo  {journal} {Nature}\ }\textbf
  {\bibinfo {volume} {600}},\ \bibinfo {pages} {75} (\bibinfo {year}
  {2021})}\BibitemShut {NoStop}%
\bibitem [{\citenamefont {Serga}\ \emph {et~al.}(2010)\citenamefont {Serga},
  \citenamefont {Chumak},\ and\ \citenamefont {Hillebrands}}]{serga2010yig}%
  \BibitemOpen
  \bibfield  {author} {\bibinfo {author} {\bibfnamefont {A.}~\bibnamefont
  {Serga}}, \bibinfo {author} {\bibfnamefont {A.}~\bibnamefont {Chumak}}, \
  and\ \bibinfo {author} {\bibfnamefont {B.}~\bibnamefont {Hillebrands}},\
  }\bibfield  {title} {\enquote {\bibinfo {title} {Yig magnonics},}\
  }\href@noop {} {\bibfield  {journal} {\bibinfo  {journal} {Journal of Physics
  D: Applied Physics}\ }\textbf {\bibinfo {volume} {43}},\ \bibinfo {pages}
  {264002} (\bibinfo {year} {2010})}\BibitemShut {NoStop}%
\bibitem [{\citenamefont {Lenk}\ \emph {et~al.}(2011)\citenamefont {Lenk},
  \citenamefont {Ulrichs}, \citenamefont {Garbs},\ and\ \citenamefont
  {M{\"u}nzenberg}}]{lenk2011building}%
  \BibitemOpen
  \bibfield  {author} {\bibinfo {author} {\bibfnamefont {B.}~\bibnamefont
  {Lenk}}, \bibinfo {author} {\bibfnamefont {H.}~\bibnamefont {Ulrichs}},
  \bibinfo {author} {\bibfnamefont {F.}~\bibnamefont {Garbs}}, \ and\ \bibinfo
  {author} {\bibfnamefont {M.}~\bibnamefont {M{\"u}nzenberg}},\ }\bibfield
  {title} {\enquote {\bibinfo {title} {The building blocks of magnonics},}\
  }\href@noop {} {\bibfield  {journal} {\bibinfo  {journal} {Physics Reports}\
  }\textbf {\bibinfo {volume} {507}},\ \bibinfo {pages} {107} (\bibinfo {year}
  {2011})}\BibitemShut {NoStop}%
\bibitem [{\citenamefont {Liu}\ \emph {et~al.}(2019)\citenamefont {Liu},
  \citenamefont {You}, \citenamefont {Wang}, \citenamefont {Xiong},\ and\
  \citenamefont {Wu}}]{liu2019phase}%
  \BibitemOpen
  \bibfield  {author} {\bibinfo {author} {\bibfnamefont {Z.-X.}\ \bibnamefont
  {Liu}}, \bibinfo {author} {\bibfnamefont {C.}~\bibnamefont {You}}, \bibinfo
  {author} {\bibfnamefont {B.}~\bibnamefont {Wang}}, \bibinfo {author}
  {\bibfnamefont {H.}~\bibnamefont {Xiong}}, \ and\ \bibinfo {author}
  {\bibfnamefont {Y.}~\bibnamefont {Wu}},\ }\bibfield  {title} {\enquote
  {\bibinfo {title} {Phase-mediated magnon chaos-order transition in cavity
  optomagnonics},}\ }\href@noop {} {\bibfield  {journal} {\bibinfo  {journal}
  {Optics Letters}\ }\textbf {\bibinfo {volume} {44}},\ \bibinfo {pages} {507}
  (\bibinfo {year} {2019})}\BibitemShut {NoStop}%
\bibitem [{\citenamefont {Wang}\ \emph
  {et~al.}(2020{\natexlab{c}})\citenamefont {Wang}, \citenamefont {Chen},
  \citenamefont {Liu}, \citenamefont {Zhang}, \citenamefont {Baumgaertl},
  \citenamefont {Guo}, \citenamefont {Li}, \citenamefont {Liu}, \citenamefont
  {Che}, \citenamefont {Tu} \emph {et~al.}}]{wang2020chiral}%
  \BibitemOpen
  \bibfield  {author} {\bibinfo {author} {\bibfnamefont {H.}~\bibnamefont
  {Wang}}, \bibinfo {author} {\bibfnamefont {J.}~\bibnamefont {Chen}}, \bibinfo
  {author} {\bibfnamefont {T.}~\bibnamefont {Liu}}, \bibinfo {author}
  {\bibfnamefont {J.}~\bibnamefont {Zhang}}, \bibinfo {author} {\bibfnamefont
  {K.}~\bibnamefont {Baumgaertl}}, \bibinfo {author} {\bibfnamefont
  {C.}~\bibnamefont {Guo}}, \bibinfo {author} {\bibfnamefont {Y.}~\bibnamefont
  {Li}}, \bibinfo {author} {\bibfnamefont {C.}~\bibnamefont {Liu}}, \bibinfo
  {author} {\bibfnamefont {P.}~\bibnamefont {Che}}, \bibinfo {author}
  {\bibfnamefont {S.}~\bibnamefont {Tu}},  \emph {et~al.},\ }\bibfield  {title}
  {\enquote {\bibinfo {title} {Chiral spin-wave velocities induced by
  all-garnet interfacial dzyaloshinskii-moriya interaction in ultrathin yttrium
  iron garnet films},}\ }\href@noop {} {\bibfield  {journal} {\bibinfo
  {journal} {Physical Review Letters}\ }\textbf {\bibinfo {volume} {124}},\
  \bibinfo {pages} {027203} (\bibinfo {year} {2020}{\natexlab{c}})}\BibitemShut
  {NoStop}%
\bibitem [{\citenamefont {Chai}\ \emph {et~al.}(2022)\citenamefont {Chai},
  \citenamefont {Shen}, \citenamefont {Zhang}, \citenamefont {Zhao},
  \citenamefont {Guo}, \citenamefont {Zou},\ and\ \citenamefont
  {Dong}}]{chai2022single}%
  \BibitemOpen
  \bibfield  {author} {\bibinfo {author} {\bibfnamefont {C.-Z.}\ \bibnamefont
  {Chai}}, \bibinfo {author} {\bibfnamefont {Z.}~\bibnamefont {Shen}}, \bibinfo
  {author} {\bibfnamefont {Y.-L.}\ \bibnamefont {Zhang}}, \bibinfo {author}
  {\bibfnamefont {H.-Q.}\ \bibnamefont {Zhao}}, \bibinfo {author}
  {\bibfnamefont {G.-C.}\ \bibnamefont {Guo}}, \bibinfo {author} {\bibfnamefont
  {C.-L.}\ \bibnamefont {Zou}}, \ and\ \bibinfo {author} {\bibfnamefont
  {C.-H.}\ \bibnamefont {Dong}},\ }\bibfield  {title} {\enquote {\bibinfo
  {title} {Single-sideband microwave-to-optical conversion in high-q
  ferrimagnetic microspheres},}\ }\href@noop {} {\bibfield  {journal} {\bibinfo
   {journal} {Photonics Research}\ }\textbf {\bibinfo {volume} {10}},\ \bibinfo
  {pages} {820} (\bibinfo {year} {2022})}\BibitemShut {NoStop}%
\bibitem [{\citenamefont {Shen}\ \emph {et~al.}(2022)\citenamefont {Shen},
  \citenamefont {Xu}, \citenamefont {Zhang}, \citenamefont {Zhang},
  \citenamefont {Wang}, \citenamefont {Chai}, \citenamefont {Zou},
  \citenamefont {Guo},\ and\ \citenamefont {Dong}}]{shen2022coherent}%
  \BibitemOpen
  \bibfield  {author} {\bibinfo {author} {\bibfnamefont {Z.}~\bibnamefont
  {Shen}}, \bibinfo {author} {\bibfnamefont {G.-T.}\ \bibnamefont {Xu}},
  \bibinfo {author} {\bibfnamefont {M.}~\bibnamefont {Zhang}}, \bibinfo
  {author} {\bibfnamefont {Y.-L.}\ \bibnamefont {Zhang}}, \bibinfo {author}
  {\bibfnamefont {Y.}~\bibnamefont {Wang}}, \bibinfo {author} {\bibfnamefont
  {C.-Z.}\ \bibnamefont {Chai}}, \bibinfo {author} {\bibfnamefont {C.-L.}\
  \bibnamefont {Zou}}, \bibinfo {author} {\bibfnamefont {G.-C.}\ \bibnamefont
  {Guo}}, \ and\ \bibinfo {author} {\bibfnamefont {C.-H.}\ \bibnamefont
  {Dong}},\ }\bibfield  {title} {\enquote {\bibinfo {title} {Coherent coupling
  between phonons, magnons, and photons},}\ }\href@noop {} {\bibfield
  {journal} {\bibinfo  {journal} {Physical Review Letters}\ }\textbf {\bibinfo
  {volume} {129}},\ \bibinfo {pages} {243601} (\bibinfo {year}
  {2022})}\BibitemShut {NoStop}%
\bibitem [{\citenamefont {Rao}\ \emph {et~al.}(2023)\citenamefont {Rao},
  \citenamefont {Yao}, \citenamefont {Wang}, \citenamefont {Zhang},
  \citenamefont {Yu},\ and\ \citenamefont {Lu}}]{rao2023unveiling}%
  \BibitemOpen
  \bibfield  {author} {\bibinfo {author} {\bibfnamefont {J.}~\bibnamefont
  {Rao}}, \bibinfo {author} {\bibfnamefont {B.}~\bibnamefont {Yao}}, \bibinfo
  {author} {\bibfnamefont {C.}~\bibnamefont {Wang}}, \bibinfo {author}
  {\bibfnamefont {C.}~\bibnamefont {Zhang}}, \bibinfo {author} {\bibfnamefont
  {T.}~\bibnamefont {Yu}}, \ and\ \bibinfo {author} {\bibfnamefont
  {W.}~\bibnamefont {Lu}},\ }\bibfield  {title} {\enquote {\bibinfo {title}
  {Unveiling a pump-induced magnon mode via its strong interaction with walker
  modes},}\ }\href@noop {} {\bibfield  {journal} {\bibinfo  {journal} {Physical
  Review Letters}\ }\textbf {\bibinfo {volume} {130}},\ \bibinfo {pages}
  {046705} (\bibinfo {year} {2023})}\BibitemShut {NoStop}%
\bibitem [{\citenamefont {Wang}\ \emph {et~al.}(2021)\citenamefont {Wang},
  \citenamefont {Yuan}, \citenamefont {Cao}, \citenamefont {Li}, \citenamefont
  {Duine},\ and\ \citenamefont {Yan}}]{wang2021magnonic}%
  \BibitemOpen
  \bibfield  {author} {\bibinfo {author} {\bibfnamefont {Z.}~\bibnamefont
  {Wang}}, \bibinfo {author} {\bibfnamefont {H.}~\bibnamefont {Yuan}}, \bibinfo
  {author} {\bibfnamefont {Y.}~\bibnamefont {Cao}}, \bibinfo {author}
  {\bibfnamefont {Z.-X.}\ \bibnamefont {Li}}, \bibinfo {author} {\bibfnamefont
  {R.~A.}\ \bibnamefont {Duine}}, \ and\ \bibinfo {author} {\bibfnamefont
  {P.}~\bibnamefont {Yan}},\ }\bibfield  {title} {\enquote {\bibinfo {title}
  {Magnonic frequency comb through nonlinear magnon-skyrmion scattering},}\
  }\href@noop {} {\bibfield  {journal} {\bibinfo  {journal} {Physical Review
  Letters}\ }\textbf {\bibinfo {volume} {127}},\ \bibinfo {pages} {037202}
  (\bibinfo {year} {2021})}\BibitemShut {NoStop}%
\bibitem [{\citenamefont {Wang}\ \emph {et~al.}(2022)\citenamefont {Wang},
  \citenamefont {Yuan}, \citenamefont {Cao},\ and\ \citenamefont
  {Yan}}]{wang2022twisted}%
  \BibitemOpen
  \bibfield  {author} {\bibinfo {author} {\bibfnamefont {Z.}~\bibnamefont
  {Wang}}, \bibinfo {author} {\bibfnamefont {H.}~\bibnamefont {Yuan}}, \bibinfo
  {author} {\bibfnamefont {Y.}~\bibnamefont {Cao}}, \ and\ \bibinfo {author}
  {\bibfnamefont {P.}~\bibnamefont {Yan}},\ }\bibfield  {title} {\enquote
  {\bibinfo {title} {Twisted magnon frequency comb and penrose
  superradiance},}\ }\href@noop {} {\bibfield  {journal} {\bibinfo  {journal}
  {Physical Review Letters}\ }\textbf {\bibinfo {volume} {129}},\ \bibinfo
  {pages} {107203} (\bibinfo {year} {2022})}\BibitemShut {NoStop}%
\bibitem [{\citenamefont {Liu}\ and\ \citenamefont
  {Li}(2022)}]{liu2022optomagnonic}%
  \BibitemOpen
  \bibfield  {author} {\bibinfo {author} {\bibfnamefont {Z.-X.}\ \bibnamefont
  {Liu}}\ and\ \bibinfo {author} {\bibfnamefont {Y.-Q.}\ \bibnamefont {Li}},\
  }\bibfield  {title} {\enquote {\bibinfo {title} {Optomagnonic frequency
  combs},}\ }\href@noop {} {\bibfield  {journal} {\bibinfo  {journal}
  {Photonics Research}\ }\textbf {\bibinfo {volume} {10}},\ \bibinfo {pages}
  {2786} (\bibinfo {year} {2022})}\BibitemShut {NoStop}%
\bibitem [{\citenamefont {Xiong}(2023)}]{xiong2023magnonic}%
  \BibitemOpen
  \bibfield  {author} {\bibinfo {author} {\bibfnamefont {H.}~\bibnamefont
  {Xiong}},\ }\bibfield  {title} {\enquote {\bibinfo {title} {Magnonic
  frequency combs based on the resonantly enhanced magnetostrictive effect},}\
  }\href@noop {} {\bibfield  {journal} {\bibinfo  {journal} {Fundamental
  Research}\ }\textbf {\bibinfo {volume} {3}},\ \bibinfo {pages} {8} (\bibinfo
  {year} {2023})}\BibitemShut {NoStop}%
\bibitem [{\citenamefont {Wan}\ \emph {et~al.}(2020)\citenamefont {Wan},
  \citenamefont {Niu}, \citenamefont {Wang}, \citenamefont {Peng},
  \citenamefont {Li}, \citenamefont {Li}, \citenamefont {Guo}, \citenamefont
  {Zou},\ and\ \citenamefont {Dong}}]{wan2020frequency}%
  \BibitemOpen
  \bibfield  {author} {\bibinfo {author} {\bibfnamefont {S.}~\bibnamefont
  {Wan}}, \bibinfo {author} {\bibfnamefont {R.}~\bibnamefont {Niu}}, \bibinfo
  {author} {\bibfnamefont {Z.-Y.}\ \bibnamefont {Wang}}, \bibinfo {author}
  {\bibfnamefont {J.-L.}\ \bibnamefont {Peng}}, \bibinfo {author}
  {\bibfnamefont {M.}~\bibnamefont {Li}}, \bibinfo {author} {\bibfnamefont
  {J.}~\bibnamefont {Li}}, \bibinfo {author} {\bibfnamefont {G.-C.}\
  \bibnamefont {Guo}}, \bibinfo {author} {\bibfnamefont {C.-L.}\ \bibnamefont
  {Zou}}, \ and\ \bibinfo {author} {\bibfnamefont {C.-H.}\ \bibnamefont
  {Dong}},\ }\bibfield  {title} {\enquote {\bibinfo {title} {Frequency
  stabilization and tuning of breathing solitons in \text{Si$_{3}$N$_{4}$}
  microresonators},}\ }\href@noop {} {\bibfield  {journal} {\bibinfo  {journal}
  {Photonics Research}\ }\textbf {\bibinfo {volume} {8}},\ \bibinfo {pages}
  {1342} (\bibinfo {year} {2020})}\BibitemShut {NoStop}%
\bibitem [{\citenamefont {Xu}\ \emph {et~al.}(2023)\citenamefont {Xu},
  \citenamefont {Zhang}, \citenamefont {Wang}, \citenamefont {Wang},
  \citenamefont {Liu}, \citenamefont {Shen}, \citenamefont {Guo},\ and\
  \citenamefont {Dong}}]{xu2023ringing}%
  \BibitemOpen
  \bibfield  {author} {\bibinfo {author} {\bibfnamefont {G.-T.}\ \bibnamefont
  {Xu}}, \bibinfo {author} {\bibfnamefont {M.}~\bibnamefont {Zhang}}, \bibinfo
  {author} {\bibfnamefont {Z.-Y.}\ \bibnamefont {Wang}}, \bibinfo {author}
  {\bibfnamefont {Y.}~\bibnamefont {Wang}}, \bibinfo {author} {\bibfnamefont
  {Y.-X.}\ \bibnamefont {Liu}}, \bibinfo {author} {\bibfnamefont
  {Z.}~\bibnamefont {Shen}}, \bibinfo {author} {\bibfnamefont {G.-C.}\
  \bibnamefont {Guo}}, \ and\ \bibinfo {author} {\bibfnamefont {C.-H.}\
  \bibnamefont {Dong}},\ }\bibfield  {title} {\enquote {\bibinfo {title}
  {Ringing spectroscopy in the magnomechanical system},}\ }\href@noop {}
  {\bibfield  {journal} {\bibinfo  {journal} {Fundamental Research}\ }\textbf
  {\bibinfo {volume} {3}},\ \bibinfo {pages} {45} (\bibinfo {year}
  {2023})}\BibitemShut {NoStop}%
\bibitem [{\citenamefont {Carmon}\ \emph {et~al.}(2004)\citenamefont {Carmon},
  \citenamefont {Yang},\ and\ \citenamefont {Vahala}}]{carmon2004dynamical}%
  \BibitemOpen
  \bibfield  {author} {\bibinfo {author} {\bibfnamefont {T.}~\bibnamefont
  {Carmon}}, \bibinfo {author} {\bibfnamefont {L.}~\bibnamefont {Yang}}, \ and\
  \bibinfo {author} {\bibfnamefont {K.~J.}\ \bibnamefont {Vahala}},\ }\bibfield
   {title} {\enquote {\bibinfo {title} {Dynamical thermal behavior and thermal
  self-stability of microcavities},}\ }\href@noop {} {\bibfield  {journal}
  {\bibinfo  {journal} {Optics Express}\ }\textbf {\bibinfo {volume} {12}},\
  \bibinfo {pages} {4742} (\bibinfo {year} {2004})}\BibitemShut {NoStop}%
\end{thebibliography}

%

\end{document}